\documentclass[aps,pra,showpacs,superscriptaddress,nofootinbib]{revtex4}
\usepackage{bbm}
\usepackage{mathrsfs}
\usepackage{epsfig,psfrag}
\usepackage{amsmath,amsfonts,amssymb}
\usepackage[usenames]{color}
\usepackage{bm}
\usepackage{ulem}
\def\mSe{ce}\def\mSo{se}
\def\mFe{Fe}\def\mFo{Fo}
\def\mJe{Je}\def\mJo{Jo}
\def\mYe{Ye}\def\mYo{Yo}

\def\mHe{He}\def\mHo{Ho}
\def\mIe{Ie}\def\mIo{Io}
\def\mKe{Ke}\def\mKo{Ko}

\begin{document}

\title{Edge Corrections to Electromagnetic Casimir Energies From
General-Purpose Mathieu Function Routines}


\author{Elizabeth Noelle Blose}
\affiliation{Department of Physics, Middlebury College, Middlebury, VT 05753, USA}
\author{Biswash Ghimire}
\affiliation{Department of Physics, Middlebury College, Middlebury, VT 05753, USA}
\author{Noah Graham}
\email{ngraham@middlebury.edu}
\affiliation{Department of Physics, Middlebury College, Middlebury, VT 05753, USA}
\author{Jeremy Stratton-Smith}
\affiliation{Department of Physics, Middlebury College, Middlebury, VT 05753, USA}

\begin{abstract}
Scattering theory methods make it possible to calculate the Casimir
energy of a perfectly conducting elliptic cylinder opposite a
perfectly conducting plane in terms of Mathieu functions.  In the
limit of zero radius, the elliptic cylinder becomes a finite-width
strip, which allows for the study of edge effects.  However,
existing packages for computing Mathieu functions are insufficient for
this calculation, because none can compute Mathieu functions of both
the first and second kind for complex arguments.  To address this
shortcoming, we have written a general purpose Mathieu function
package, based on algorithms developed by Alhargan
\cite{Alhargan:2000,Alhargan:2000a}.  We use these routines to
find edge corrections to the proximity force approximation for the
Casimir energy of a perfectly conducting strip opposite a
perfectly conducting plane.
\end{abstract}

\maketitle

\section{Introduction}
Scattering theory methods have made it possible to calculate
Casimir energies, arising from quantum-mechanical fluctuations of
charges and fields in quantum electrodynamics, for any objects for
which one can obtain the $T$-matrices for light scattering.  In this
approach, one expresses the Casimir energy in ``TGTG'' form
\cite{Kenneth06}, generalizing scattering results for planar
geometries \cite{Lambrecht06} and for scalar fields in spherical
geometries \cite{Bulgac06} to any geometry that is tractable for
electromagnetic scattering \cite{spheres,universal}.  The resulting
calculation combines the scattering $T$-matrix, which captures the
reflection of quantum fluctuations from each object individually, with
the free Green's function, which propagates fluctuations from one
object to the other.

One geometry of particular interest is the electromagnetic cylinder,
which in the limit of zero radius becomes a finite-width strip
\cite{Graham:2013rga}, allowing for the study of edge effects
\cite{Graham:2013rga,wedge,parabolic1,parabolic2,Gies3,Gies4,Kabat1,Kabat2}.
These effects can be modeled as corrections to the proximity force
approximation (for a situation where the derivative expansion
\cite{Fosco:2011xx,beyondpfa,Teo:2013kwa} does not apply).  However,
numerical calculations of the Casimir energy in elliptic cylinder
geometries require Mathieu functions
\cite{McLachlan:104546,Abramowitz,Morse53,Bateman}, in cases
for which existing packages are not well-suited.  As a result, we have
created a general-purpose package to compute odd and even, angular and
radial, first-kind and second-kind, and ordinary and modified Mathieu
functions of integer order, for complex parameter and argument and
integer index.  Our approach is based on the routines developed by
Alhargan \cite{Alhargan:2000,Alhargan:2000a}, extended to the case of
complex inputs and implemented in {\sc Mathematica}.

At short distances, we can expand the Casimir interaction energy per
unit length of a perfectly conducting strip oriented parallel to a
perfectly conducting plane as
\begin{equation}
\frac{\cal E}{\hbar c L} = -\frac{\pi^2}{720} \frac{2d}{H^3}
+ \frac{2 \beta}{H^2} + \frac{\gamma}{2d H} + \ldots
\label{eqn:expansion}
\end{equation}
where $2d$ is the width of the strip, $H$ is the distance between the
plane and the strip, and $\beta$ and $\gamma$ are dimensionless
constants.  The leading term in this expansion gives the proximity force
approximation; the second term gives the interaction of the two edges
with the infinite plane; and the third term gives the interaction
between the two edges, mediated by the plane.  From an exact numerical
calculation, we find good agreement with this form, with
$\beta=0.00092$ and $\gamma=-0.0040$.  The former quantity agrees with
results obtained at lower precision in 
Refs.~\cite{wedge,parabolic1,parabolic2,planes}; its small magnitude 
can be explained by the cancellation of the effects of the first
reflection for electromagnetism \cite{wedge,Babinet,planes}.

In the following sections we assemble the various components needed to
obtain this result.  In Section \ref{sec:scattering} we review
scattering theory in elliptic cylinder coordinates and establish
conventions for the Mathieu functions that arise as solutions to the
Helmholtz equation.  Then in Section \ref{sec:Mathieu} we describe the
numerical package we have developed to calculate these functions in
the generality required for Casimir calculations.  Finally, we discuss
the results of the Casimir calculation in Section \ref{sec:discussion}.

\section{Scattering Theory in Elliptic Cylinder coordinates}
\label{sec:scattering}

We begin by formulating scattering theory in elliptic cylinder
coordinates,
\begin{equation}
 x = d\cosh \mu\cos \theta \qquad
 y = d\sinh \mu\sin \theta \qquad
 z = z \,,
\end{equation}
where $2d$ is the interfocal separation, $-\pi < \theta \leq \pi$ is the
analog of the angle in ordinary cylindrical coordinates, and $0\leq
\mu < \infty$ is the analog of the ordinary cylindrical radius $R$, with 
\begin{equation}
R = \sqrt{ x^2 +  y^2} = 
d\sqrt{\frac{\cosh 2 \mu+\cos 2 \theta}{2}}\to \frac{d}{2} e^{ \mu} 
\hbox{\qquad as $ \mu\to\infty$.}
\end{equation}

The Helmholtz equation in elliptic cylinder coordinates is given by
\begin{equation}
\frac{1}{d^2\left(\cosh^2\mu - \cos^2\theta \right)}
\left(\frac{\partial^2\Psi}{\partial \mu^2}+
\frac{\partial^2\Psi}{\partial \theta^2}\right) +
\frac{\partial^2\Psi}{\partial z^2} + k^2 \Psi = 0 \,.
\end{equation}
Using separation of variables with $\Psi = M(\mu) \Theta(\theta) Z(z)$ gives 
\begin{eqnarray}
\frac{d^2 \Theta}{d\theta^2}+ \left(\alpha - 2q \cos 2\theta
\right)\Theta(\theta) &=& 0\cr
\frac{d^2 M}{d \mu^2}- \left(\alpha - 2q \cosh \mu \right)M(\mu) &=& 0 \cr
\frac{d^2 Z}{d z^2} + k_z^2 Z(z) &=& 0 \,,
\end{eqnarray}
where the parameter $q$ is given by $q=\frac{d^2}{4}(k^2-k_z^2)$ and
the separation constant $\alpha$ is  known as the characteristic
value.  For $Z(z)$ we have the standard complex exponential solutions 
$Z(z) = e^{i k_z z}$. Because the problem still has reflection
symmetry, we have angular solutions $\Theta(\theta)$
that are either even or odd functions of the
argument $\theta$, the analogs of $\cos$ and $\sin$ in the ordinary
cylinder case.  We are interested only in characteristic values
for which the resulting angular functions are periodic, which we label
by the integer index $r$, where $r$ runs from $0$ to $\infty$
for even solutions and from $1$ to $\infty$ for odd solutions.  For
$q=0$ and $r \neq 0$, the even and odd angular functions then reduce to
$\cos r\theta$ and $\sin r\theta$ respectively, and the characteristic
value becomes $r^2$ in both cases.  For the special case of $r=0$, in the
limit as $q$ goes to zero the even angular function goes not to $\cos
0 =1$, but instead to the constant function $\frac{1}{\sqrt{2}}$, with
characteristic value zero (and there is no odd angular function for $r=0$).

The radial solutions $M(\mu)$ are the analogs
of Bessel functions in ordinary cylindrical coordinates.
We note that the radial functions obey the same differential
equation as the angular functions with imaginary argument, a
relationship we will make use of in our computational algorithm.
Unlike the case of ordinary cylindrical functions, the radial functions
corresponding to even and odd angular functions for the same index $r$
differ, because they have different characteristic values.

Because the Mathieu equations are second-order, they each have two
independent solutions:  solutions of the first kind obey appropriate
regularity conditions at the origin, while solutions of the second
kind do not.  Furthermore, since $q=\frac{d^2}{4}(k^2-k_z^2)$, positive
values of $q$ correspond to propagating waves, while negative values
of $q$ correspond to evanescent waves; it will be convenient to
define modified versions of all of our functions, which are related to
the ordinary functions with $q\to -q$.  These choices --- even and odd,
angular and radial, ordinary and modified, and first-kind and
second-kind --- therefore yield a total 16 Mathieu functions.  The
four modified angular functions typically are not assigned their own
names; the remaining 12 are summarized in Table~\ref{table:functions}.

\begin{table}
\begin{tabular}{|c|c|c|c|c|c|c|c|c|c|c|c|}
\hline
\multicolumn{4}{|c|}{Angular} &
\multicolumn{8}{|c|}{Radial} \\ \hline
\multicolumn{4}{|c|}{Ordinary}& 
\multicolumn{4}{|c|}{Ordinary}  & \multicolumn{4}{c|}{Modified} \\ \hline
\multicolumn{2}{|c|}{First kind} & 
\multicolumn{2}{c|}{Second kind} & \multicolumn{2}{c|}{First kind} &
\multicolumn{2}{|c|}{Second kind} & 
\multicolumn{2}{c|}{First kind} & \multicolumn{2}{c|}{Third kind} \\ \hline
Even & Odd & Even & Odd & Even & Odd & Even & Odd &
Even & Odd & Even & Odd  \\ \hline
$\mSe_r(q,\theta)$ & $\mSo_r(q,\theta)$ & $\mFe_r(q,\theta)$ & $\mFo_r(q,\theta)$ &
$\mJe_r(q,\mu)$ & $\mJo_r(q,\mu)$ & $\mYe_r(q,\mu)$ & $\mYo_r(q,\mu)$ & 
$\mIe_r(q,\mu)$ & $\mIo_r(q,\mu)$ & $\mKe_r(q,\mu)$ & $\mKo_r(q,\mu)$ \\ \hline
\end{tabular}
\caption{Table of Mathieu functions.  Modified angular functions are
not assigned separate names; they are simply given by sending $q\to
-q$ in the ordinary functions.  Note that the modified
functions $\mKe$ and $\mKo$ are referred to as ``third kind''
because they are related not to $\mYe$ and $\mYo$, but
instead to the combinations $\mHe = \mJe + i \mYe$ and 
$\mHo = \mJo + i \mYo$, as described below.
}
\label{table:functions}
\end{table}

We normalize our functions following the conventions of
of Abramowitz and Stegun \cite{Abramowitz}, but name them using a 
modified notation that is more closely analogous to the ordinary
cylinder case.  The even and odd angular functions 
$\mSe_r(q,\theta)$ and $\mSo_r(q,\theta)$ are normalized such that
\begin{equation}
\int_0^{2\pi} \mSe_r(q,\theta)^2 d\theta  = \int_0^{2\pi}
\mSo_r(q,\theta)^2 d\theta = \pi \,,
\end{equation}
which is analogous to the normalization of the ordinary trigonometric
functions $\cos r \theta$ and $\sin r \theta$ (except for $r=0$, as
described above). The radial functions are all normalized so that they
approach the analogous Bessel functions at large distances.  
These conventions are convenient for creating a standard expansion
of free quantum fields in terms of Mathieu functions \cite{Graham:2005cq}.
Also in analogy with Bessel functions, we define the modified functions by
\begin{equation}
\mIe_r(-q,\mu) = i^{-r} \mJe_r(q,\mu) \qquad
\mIo_r(-q,\mu) = i^{-r} \mJo_r(q,\mu)
\end{equation}
and
\begin{equation}
\mKe_r(-q,\mu) = i^{r+1} \frac{\pi}{2} \mHe_r(q,\mu) \qquad
\mKo_r(-q,\mu) = i^{r+1} \frac{\pi}{2} \mHo_r(q,\mu)
\end{equation}
for $q<0$, where the radial functions of the third kind are given by
$\mHe_r(q,\mu) = \mJe_r(q,\mu) + i \mYe_r(q,\mu)$ and
$\mHo_r(q,\mu) = \mJo_r(q,\mu) + i \mYo_r(q,\mu)$.  
As in the ordinary cylinder case, the definitions
of $\mKe$ and $\mKo$ in terms of third kind functions yield the
exponentially decaying evanescent solutions, avoiding the cancellation of
large numbers that would be required to extract these solutions from the
direct continuation to negative $q$ of the solutions first and second kind.

\section{Computation of Mathieu Functions}
\label{sec:Mathieu}

We have developed a package written in {\sc
Mathematica} for computing Mathieu functions.  The
built-in functionality of {\sc Mathematica} supports only angular,
first-kind functions (similar functionality is available in {\sc
Maple}).  Since complex arguments are allowed, one can in principle
obtain the radial first-kind functions as well.  However, as described
below, the standard calculation of angular functions is of limited
utility for arguments with nonzero imaginary part; as a result, in
that case we will need to use routines designed explicitly for the
calculation of radial functions.  We will also need second-kind
functions to describe irregular scattering waves.

Our approach is based on methods developed by Alhargan
\cite{Alhargan:2000,Alhargan:2000a}.  While these routines, written in
C, support all 16 Mathieu functions and work reliably for all the
inputs we have tested, they accept only positive $q$ and real
arguments.  We will therefore generalize these routines to
complex argument and parameter, motivated by the Casimir
calculation, which involves both angular functions of the first kind
with complex argument and modified radial functions of the first and
third kinds.  Our code is available at
{\tt http://community.middlebury.edu/\~{}ngraham}\,.

\subsection{Angular Functions for Real Argument}

The standard calculation of angular Mathieu functions uses a Fourier
series expansion,
\begin{eqnarray}
\mSe_r\left(q,\theta \right) &=& 
\frac{\delta }{\sqrt{\sum\limits_{m=0}^\infty
A_{2m+p}\left(r,q\right)^2+ (1-p)}}
\sum_{m=0}^\infty A_{2m+p}\left(r,q\right)
\cos \left[\left(2m+p\right)\theta\right] \cr
\hbox{with} &&
    p = 
        \begin{cases}
         1, &   \text{odd } r  \\
         0, &  \text{even } r
        \end{cases}
\hbox{\quad and \quad}
\delta = 
        \begin{cases}
         \left(-1\right)^{\left(r-p\right)/2}, & \text{for } \Re\left(q\right) < 0  \\
         1, & \text{otherwise}
        \end{cases} \,,
\label{eqn:angularseries1}
\end{eqnarray}
and 
\begin{eqnarray}
\mSo_r\left(q,\theta \right) &=&
\frac{\delta}{\sqrt{\sum\limits_{m=0}^\infty
B_{2m+p}\left(r,q \right)^2}}\sum_{m=0}^\infty B_{2m+p}\left(r,q\right) 
\sin \left[\left(2m+p\right) \theta\right] \cr
\hbox{with} &&
    p = 
        \begin{cases}
         1, &   \text{odd } r  \\
         0, &  \text{even } r
        \end{cases}
\hbox{\quad and \quad}
\delta = 
        \begin{cases}
         \left(-1\right)^{\left(r-2+p\right)/2}, & \text{for } \Re\left(q\right) < 0  \\
         1, & \text{otherwise}
        \end{cases}
\label{eqn:angularseries2}
\end{eqnarray}
Here the prefactors implement our $L^2$ normalization convention for
the angular functions (which differs from that used by Alhargan).  To
obtain the coefficients $A_{2m+p}\left(r,q\right)$ and
$B_{2m+p}\left(r,q\right)$, we follow
Alhargan and use both upward and downward recurrence relations
for the ratios of adjacent coefficients.  These recurrences start from
zero and infinity respectively, and then meet at $m=r$.  The
forms are slightly different for the even and odd functions and for
odd and even order $r$.

For the even function coefficients, denoting the even characteristic
values as $a_r$ we have the following recursions,
\begin{eqnarray}
\textbf{Even order}& \qquad &\textbf{Odd order}\cr
A_m\left(r,q\right)= Ae_m= Ae_{m-2}Ve_{m-2},& \qquad &A_m\left(r,q\right) = Ao_m = Ao_{m-2} Vo_{m-2}
\end{eqnarray}
for $m>1$ with the base cases 
\begin{eqnarray}
Ae_0 = 1 & \qquad & Ao_1 = 1 \cr
Ve_0 = \frac{a_r}{q}& \qquad &Vo_1 = -1 +\frac{a_{r}-1}{q} \cr
Ve_2 = \frac{a_r -4}{q}-\frac{2}{Ve_0}& \qquad & \cr
Ve_{\infty} = 0&\qquad & Vo_{\infty} = 0
\end{eqnarray}
and the recursion relations for $m>2$
\begin{eqnarray}
    Ve_m = \begin{cases}
      \displaystyle
         \frac{a_r - m^2}{q}-\frac{1}{Ve_{m-2}}, &   m\leq r  \\
         \frac{-q}{\left(m+2\right)^2 - a_r +q Ve_{m+2}}, &   m>r
        \end{cases}
& \qquad &
    Vo_m =\begin{cases}
      \displaystyle
         \frac{a_r - m^2}{q}-\frac{1}{Vo_{m-2}}, &   m\leq r  \\
         \frac{-q}{\left(m+2\right)^2 - a_r +q Vo_{m+2}}, &   m>r
        \end{cases} \,.
\end{eqnarray}
For the odd function coefficients, denoting the odd
characteristic values as $b_r$ we have
\begin{eqnarray} 
\textbf{Even order}& \qquad &\textbf{Odd order} \cr
B_m\left(r,q\right)=Be_m=Be_{m-2}We_{m-2},&
&B_m\left(r,q\right)=Bo_m=Bo_{m-2}Wo_{m-2}
\end{eqnarray}
for $m>2$ with the base cases
\begin{eqnarray}
Be_0=We_0 = 0 & \qquad & \cr
Be_2 = 1 & \qquad & Bo_1 = 1 \cr
We_2 =\frac{-4+b_r}{2}&\qquad & 
Wo_1 = 1+\frac{b_r -1}{q} \cr
Be_{\infty} = 0,& &Bo_{\infty} = 0
\end{eqnarray}
and the recursion relations for $m>2$
\begin{eqnarray}
    We_m = \begin{cases}
      \displaystyle
         \frac{b_r-m^2}{q}+\frac{-1}{We_{m-2}}, &   m\leq r  \\
         \frac{-q}{\left(m+2\right)^2 - b_r +q We_{m+2}}, &   m>r
        \end{cases}
& \qquad &
    Wo_m =\begin{cases}
      \displaystyle
         \frac{b_r - m^2}{q}-\frac{1}{Wo_{m-2}}, &   m\leq r  \\
         \frac{-q}{\left(m+2\right)^2 - b_r +q Wo_{m+2}}, &   m>r
        \end{cases} \,.
\end{eqnarray}

\subsection{Radial Functions}

Again following Alhargan \cite{Alhargan:2000,Alhargan:2000a}, we find
the radial functions as expansions in products of Bessel functions, in
terms of the same coefficients as we found in the angular case.  These
expansions take the form
\begin{eqnarray}
\mJe_r\left(q,\mu\right) &=& \frac{\sigma_r (-1)^{\frac{r-p}{2}}}{A_r\left(r,q\right)}\sum_{m=0}^\infty \left(-1\right)^{m} A_{2m+p}\left(r,q\right)
\left[J_{m-\frac{r-p}{2}}\left(e^{-\mu}\sqrt{q}\right)J_{m+\frac{r+p}{2}}\left(e^\mu\sqrt{q}\right)
\right. \cr && \qquad \qquad \qquad \qquad \qquad \qquad \qquad + \left.
J_{m+\frac{r+p}{2}}\left(e^{-\mu}\sqrt{q}\right)J_{m-\frac{r-p}{2}}\left(e^\mu
\sqrt{q}\right)\right]  \cr
\mJo_r\left(q,\mu \right) &=&\frac{(-1)^{\frac{r-p}{2}}}{B_r\left(r,q\right)}\sum_{m=0}^\infty \left(-1\right)^{m} B_{2m+p}\left(r,q\right)
\left[J_{m-\frac{r-p}{2}}\left(e^{-\mu}\sqrt{q}\right)J_{m+\frac{r+p}{2}}\left(e^\mu\sqrt{q}\right)
\right. \cr && \qquad \qquad \qquad \qquad \qquad \qquad \qquad - \left.
J_{m+\frac{r+p}{2}}\left(e^{-\mu}\sqrt{q}\right)J_{m-\frac{r-p}{2}}\left(e^\mu
\sqrt{q}\right)\right] \cr
\mYe_r\left(q,\mu \right) &=&\frac{\sigma_r (-1)^{\frac{r-p}{2}}}{A_r\left(r,q\right)}\sum_{m=0}^\infty \left(-1\right)^{m} A_{2m+p}\left(r,q\right)
\left[J_{m-\frac{r-p}{2}}\left(e^{-\mu}\sqrt{q}\right)Y_{m+\frac{r+p}{2}}\left(e^\mu\sqrt{q}\right)
\right. \cr && \qquad \qquad \qquad \qquad \qquad \qquad \qquad + \left.
J_{m+\frac{r+p}{2}}\left(e^{-\mu}\sqrt{q}\right)Y_{m-\frac{r-p}{2}}\left(e^\mu
\sqrt{q}\right)\right] \cr
\mYo_r\left(q,\mu \right) &=& \frac{(-1)^{\frac{r-p}{2}}}{B_r\left(r,q\right)}\sum_{m=0}^\infty \left(-1\right)^{m} B_{2m+p}\left(r,q\right)
\left[J_{m-\frac{r-p}{2}}\left(e^{-\mu}\sqrt{q}\right)Y_{m+\frac{r+p}{2}}\left(e^\mu\sqrt{q}\right)
\right. \cr && \qquad \qquad \qquad \qquad \qquad \qquad \qquad - \left.
J_{m+\frac{r+p}{2}}\left(e^{-\mu}\sqrt{q}\right)Y_{m-\frac{r-p}{2}}
\left(e^\mu \sqrt{q}\right)\right] \cr
\mIe_r\left(q,\mu\right) &=& \frac{\sigma_r}{A_r\left(r,q\right)}\sum_{m=0}^\infty A_{2m+p}\left(r,q\right)
\left[I_{m-\frac{r-p}{2}}\left(e^{-\mu}\sqrt{q}\right)I_{m+\frac{r+p}{2}}\left(e^\mu\sqrt{q}\right)
\right. \cr && \qquad \qquad \qquad \qquad \qquad \qquad \qquad + \left.
I_{m+\frac{r+p}{2}}\left(e^{-\mu}\sqrt{q}\right)I_{m-\frac{r-p}{2}}\left(e^\mu
\sqrt{q}\right)\right] \cr
\mIo_r\left(q,\mu\right) &=& \frac{1}{B_r\left(r,q\right)}\sum_{m=0}^\infty B_{2m+p}\left(r,q\right)
\left[I_{m-\frac{r-p}{2}}\left(e^{-\mu}\sqrt{q}\right)I_{m+\frac{r+p}{2}}\left(e^\mu\sqrt{q}\right)
\right. \cr && \qquad \qquad \qquad \qquad \qquad \qquad \qquad - \left.
I_{m+\frac{r+p}{2}}\left(e^{-\mu}\sqrt{q}\right)I_{m-\frac{r-p}{2}}\left(e^\mu
\sqrt{q}\right)\right] \cr
\mKe_r\left(q,\mu\right) &=& \frac{\sigma_r\left(-1\right)^{\frac{r-p}{2}}}{A_r\left(r,q\right)}\sum_{m=0}^\infty\left(-1\right)^m A_{2m+p}\left(r,q\right)
\left[I_{m-\frac{r-p}{2}}\left(e^{-\mu}\sqrt{q}\right)K_{m+\frac{r+p}{2}}\left(e^\mu\sqrt{q}\right) 
\right. \cr && \qquad \qquad \qquad \qquad \qquad \qquad \qquad + \left.
\left(-1\right)^pI_{m+\frac{r+p}{2}}\left(e^{-\mu}\sqrt{q}\right)K_{m+\frac{r-p}{2}}\left(e^\mu
\sqrt{q}\right)\right]  \cr
\mKo_r\left(q,\mu\right) &=&
\frac{\left(-1\right)^{\frac{r-p}{2}}}{B_r\left(r,q\right)}\sum_{m=0}^\infty\left(-1\right)^m B_{2m+p}\left(r,q\right)
\left[I_{m-\frac{r-p}{2}}\left(e^{-\mu}\sqrt{q}\right)K_{m+\frac{r+p}{2}}\left(e^\mu
\sqrt{q}\right)
\right. \cr && \qquad \qquad \qquad \qquad \qquad \qquad \qquad - \left.
\left(-1\right)^pI_{m+\frac{r+p}{2}}\left(e^{-\mu}\sqrt{q}\right)K_{m+\frac{r-p}{2}}\left(e^\mu \sqrt{q}\right)\right]
\end{eqnarray}
with
\begin{equation}
    p = 
        \begin{cases}
         1, &   \text{odd } r  \\
         0, &  \text{even } r
        \end{cases}
\hbox{\qquad and \qquad}
    \sigma_{r} =
        \begin{cases}
        \frac{1}{2}, & r = 0 \\
        1, & r \neq 0
        \end{cases} \,.
\end{equation}

\subsection{Angular Functions for Complex Argument}

For complex arguments, the Fourier series in
Eqs.~(\ref{eqn:angularseries1}) and (\ref{eqn:angularseries2}) 
become numerically ill-behaved.  This
problem does not, however, affect the Bessel function series used to
compute the radial functions.  Since the radial functions obey the
same differential equation as the angular functions of imaginary
argument (and vice versa), these functions differ only by a
normalization factor.  We take advantage this relationship to write
\begin{equation}
\mSe_r\left(q,\theta\right)=
\frac{\mSe_r\left(q,0\right)}{\mJe_r\left(q,0\right)}
\mJe_r\left(q,-i\theta\right)
\qquad
\mSo_r\left(q,\theta\right)=
\frac{\mSo_r'\left(q,0\right)}{\mJo_r'\left(q,0\right)}
\mJo_r\left(q,-i\theta\right) \,,
\end{equation}
where prime denotes a derivative with respect to the argument.  The
prefactor ratios in both expressions, which are independent of
argument, serve as ``joining factors'' to convert the normalizations
of the two functions.  We therefore use these relationships, along with our
routines for radial functions, to compute the angular functions of
complex argument.  We also use this approach any time the magnitude of
$q$ is very small, again to avoid numerical instabilities.  
While we use Eqs.~(\ref{eqn:angularseries1}) and
(\ref{eqn:angularseries2}) for the case of real argument, the
corresponding radial function expansions would also work perfectly
well, but they are slower to compute.

\subsection{Second Kind Angular Functions}

Although they are not needed in the Casimir calculation, for
completeness our code also implements angular functions of the second
kind, again using the approach of Alhargan
\cite{Alhargan:2000,Alhargan:2000a}.  These functions can be written
as
\begin{equation}
\mFe_r\left(q, \theta \right) =
\frac{\frac{2\delta\sqrt{\alpha_2 + 1 - p}}
{\pi(1+\alpha_1)}}{\alpha_1\left(1+
\sum\limits_{m=0}^\infty
\frac{\left(2m+p\right) G_{2m+p}\left(r,q\right)}{\alpha_1}\right)}
\left( \theta \sqrt{\alpha_2}  \mSe_r\left(q, \theta \right) +
\sum_{m=0}^\infty G_{2m+p}\left(r,q\right)
\sin \left[\left(2m+p\right)\theta \right]\right)
\end{equation}
with
\begin{equation}
\alpha_i = \sum_{m=0}^\infty A_{2m+p}\left(r,q\right)^i\,,
\quad
    p = 
        \begin{cases}
         1, &   \text{odd } r  \\
         0, &  \text{even } r
        \end{cases} \,,
\hbox{\quad and \quad}
    \delta = 
        \begin{cases}
         \left(-1\right)^{\left(r-p\right)/2}, & \text{for } \Re(q) < 0  \\
         1, & \text{otherwise}
        \end{cases} \,.
\end{equation}
for the even functions and
\begin{equation}
\mFo_r\left(q, \theta \right) =
\frac{\frac{2\delta\sqrt{\alpha_{2,0}}}{\pi(2-p+\alpha_{1,1})}}
{\left(2-p + \alpha_{1,1}\right)
+ \sum\limits_{m=0}^\infty
\frac{H_{2m+p}\left(r,q\right)}{\alpha_{1,1}}}
\left(\theta \sqrt{\alpha_{2,0}}
\mSo_r\left(q, \theta \right) +
\sum_{m=0}^\infty H_{2m+p}\left(r,q\right) 
\cos \left[\left(2m+p\right)\theta \right]\right)
\end{equation}
with
\begin{equation}
\alpha_{i,j} = \sum_{m=0}^\infty \left(2m+p\right)^j
B_{2m+p}\left(r,q\right)^i \,,
\quad
    p = 
        \begin{cases}
         1, &   \text{odd } r  \\
         0, &  \text{even } r
        \end{cases} \,,
\hbox{\quad and \quad}
    \delta = 
        \begin{cases}
         \left(-1\right)^{\left(r-2+p\right)/2}, & \text{for } \Re(q) < 0  \\
         1, & \text{otherwise}
        \end{cases}
\end{equation}
for the odd functions.  Similarly to the first-kind case, we compute the even coefficients via
\begin{eqnarray}
\textbf{Even order recursion}& \qquad &\textbf{Odd order recursion}\cr
Ge_n\left(r,q\right) = Ge_n =Qe_n - \rho_{qe} Ae_n && 
Go_n\left(r,q\right) = Go_n = Qo_n - \rho_{qo} Ao_n \cr
\rho_{qe}
=\frac{1}{2Ae_0}\left[\frac{\left(a_r-4\right)Qe_2}{q}-Qe_4\right]
-\frac{2a_r}{q^2} 
&&
\rho_{qo} =
\frac{1}{2Ao_1}\left[\frac{\left(a_r-1+q\right)Qo_1}{q}-Qo_3\right]
-\frac{1}{q} 
\end{eqnarray}
with base cases
\begin{eqnarray}
Qe_{2n_{\rm max}}=0 &\qquad& Qo_{2n_{\rm max}+1}=0 \cr
Qe_{2n_{\rm max}-2} =-\frac{4n_{\rm max}~Ae_{2n_{\rm max}} }{q} &&
Qo_{2n_{\rm max}-1}=-\frac{2\left(2n_{\rm max}+1\right)Ao_{2n_{\rm max}+1}}{q}
\end{eqnarray}
and recursion relations
\begin{eqnarray}
Qe_{n-2} = \frac{\left(a_r - n^2\right) Qe_{n}-2 n Ae_{n}}{q}-Qe_{n+2}
&&
Qo_{n-2} = \frac{\left(a_r - n^2\right) Qo_{n}-2 n Ao_{n}}{q}-Qo_{n+2}
\end{eqnarray}
For the odd coefficients, we have
\begin{eqnarray}
\textbf{Even order recursion}& \qquad &\textbf{Odd order recursion}\cr
He_n(r,q) = He_n =Te_n - \rho_{te} Be_n && Ho_n(r,q) = 
Ho_n = To_n - \rho_{to} Bo_n \cr
\rho_{te} =\frac{1}{Be_2}\left[Te_2-\frac{b_r Te_0}{q}\right]&
&\rho_{to} = \frac{1}{2Bo_1}\left[To_3 -
\frac{\left(b_r-1-q\right)To_1}{q}\right]-\frac{1}{q}
\end{eqnarray}
with base cases
\begin{eqnarray}
Te_{2n_{\rm max}}=0 &\qquad& To_{2n_{\rm max}+1} =0 \cr
Te_{2n_{\rm max}-2} =-\frac{4n_{\rm max}~Be_{2n_{\rm max}} }{q} &&
To_{2n_{\rm max}-1}=\frac{2\left(2n_{\rm max}+1\right)Bo_{2n_{\rm max}+1}}{q}
\end{eqnarray}
and recursion relations
\begin{eqnarray}
Te_{n-2} = \frac{\left(b_r - n^2\right) Te_{n}+2 n Be_{n}}{q}-Te_{n+2}
&&
To_{n-2} = \frac{\left(b_r - n^2\right) To_{n}+2 n Bo_{n}}{q}-To_{n+2} \,.
\end{eqnarray}

\subsection{Implementation Details}

We note a number of design elements of our code, which serve to
enhance its efficiency, convenience, and reliability.
\begin{itemize}
\item
Characteristic values are computed using the built-in functions in
{\sc Mathematica}.
\item
Since the Mathieu functions solve second-order differential equations,
it is helpful to have expressions for their first derivatives with
respect to their arguments.  We implement these by differentiating the
corresponding series expansions term by term, which we can then
simplify using known properties of derivatives of trigonometric and
Bessel functions.  
\item
The Wronskian relations for the first- and second-kind functions
and their first derivatives provide valuable checks on the numerical
calculation.
\item
Quantities that are likely to be needed repeatedly, such as joining
factors and coefficients in recurrence relations, are cached.
\item
For the case of radial functions with real arguments, stable
recurrence relations are used to efficiently compute Bessel functions
for the entire range of orders needed.
\end{itemize}

\section{Casimir Calculation and Discussion}
\label{sec:discussion}

As shown in Ref.~\cite{Graham:2013rga}, the energy per unit length of
a perfectly reflecting strip opposite a perfectly reflecting plane
is given in terms of the angular Mathieu functions $\mSe_r$ and $\mSo_r$
by
\begin{equation}
\frac{\cal E}{\hbar c L}=
\frac{1}{4\pi}\int_0^\infty p dp
\log \det \left[\mathbbm{1}_{rr'}^{\chi \chi'} - 
{\cal T}^\chi_{r} {\cal T}^P
\int_{-\infty}^\infty du \, e^{-2 p H \cosh u}
\genfrac{}{}{0pt}{}{\mSe_r}{\mSo_r}
\left(q,\frac{\pi}{2} + iu+ \varphi\right)
\genfrac{}{}{0pt}{}{\mSe_{r'}}{\mSo_{r'}}
\left(q,\frac{\pi}{2} - iu + \varphi\right)
\right]\,,
\label{eqn:energy}
\end{equation}
where $\chi$ and $\chi'$ denote the odd and even scattering channels,
corresponding to $\mSe_r$ and $\mSo_r$ respectively, $H$ is the height
of the center of strip above the plane, $q=-\frac{d^2 p^2}{4}$,
$\varphi$ is the angle of the strip relative to the plane,
and the determinant runs over the $r$
and $r'$ indices and both parity channels.  The scattering $T$-matrix
for the plane is given by ${\cal T}^P = \pm 1$  for Neumann and
Dirichlet boundary conditions respectively, while for an elliptic
cylinder of radius $\mu_0$ we have $\displaystyle {\cal T}_{r k_z r'
  k_z'}^{e,o} =  2\pi \delta(k_z - k_z')\delta_{rr'} {\cal
  T}_r^{e,o}$, with
\begin{eqnarray}
{\cal T}_r^e =
-\frac{\mIe_r \left(-q,\mu_0 \right)} {\mKe_r\left(-q,\mu_0 \right)} \qquad
{\cal T}_r^o =
-\frac{\mIo_r \left(-q,\mu_0 \right)} {\mKo_r\left(-q,\mu_0 \right)}
&& \hbox{\qquad (Dirichlet)} \cr
{\cal T}_r^e =
-\frac{\mIe_r' \left(-q,\mu_0 \right)} {\mKe_r' \left(-q,\mu_0 \right)} \qquad
{\cal T}_r^o =
-\frac{\mIo_r' \left(-q,\mu_0 \right)} {\mKo_r' \left(-q,\mu_0 \right)}
&& \hbox{\qquad (Neumann)}
\end{eqnarray}
for our two boundary conditions, where prime indicates a derivative
with respect to $\mu$.  To obtain the Casimir energy for
electromagnetism with perfect conductors, we take the sum of this
result for Dirichlet and Neumann boundary conditions (with the same
boundary condition on both surfaces).  The case of the strip is then
given by taking $\mu_0 = 0$ in these results.  We will consider
$\varphi=0$, so that the strip is parallel to the plane.  In that case
we can simplify Eq.~(\ref{eqn:energy}) via the identities
\begin{eqnarray}
\mSe_r(q,\theta) &=& 
\left\{ \begin{array}{l@{\quad}l}
(-1)^{\frac{r}{2}} \mSe_r\left(-q,\frac{\pi}{2} - \theta\right) &
\hbox{for $r$ even} \cr
(-1)^{\frac{r-1}{2}} \mSo_r\left(-q,\frac{\pi}{2} - \theta\right) &
 \hbox{for $r$ odd}
\end{array} \right. \cr
\mSo_r(q,\theta) &=& 
\left\{ \begin{array}{l@{\quad}l}
(-1)^{\frac{r}{2}-1} \mSo_r\left(-q,\frac{\pi}{2} - \theta\right) &
\hbox{for $r$ even} \cr
(-1)^{\frac{r-1}{2}} \mSe_r\left(-q,\frac{\pi}{2} - \theta\right) &
 \hbox{for $r$ odd}
\end{array} \right. \,,
\end{eqnarray}
so that for $\varphi=0$, we require the angular Mathieu functions at
purely imaginary argument.  For $\varphi=0$, the integrand in
Eq.~(\ref{eqn:energy}) is also a symmetric or antisymmetric function
of $u$, and the determinant decomposes into two independent sectors,
one consisting of the modes for which the parity of the elliptic
functions matches the parity of $r$, and the other the modes for which
the parities are opposite.

\begin{figure}
\includegraphics[width=0.6\linewidth]{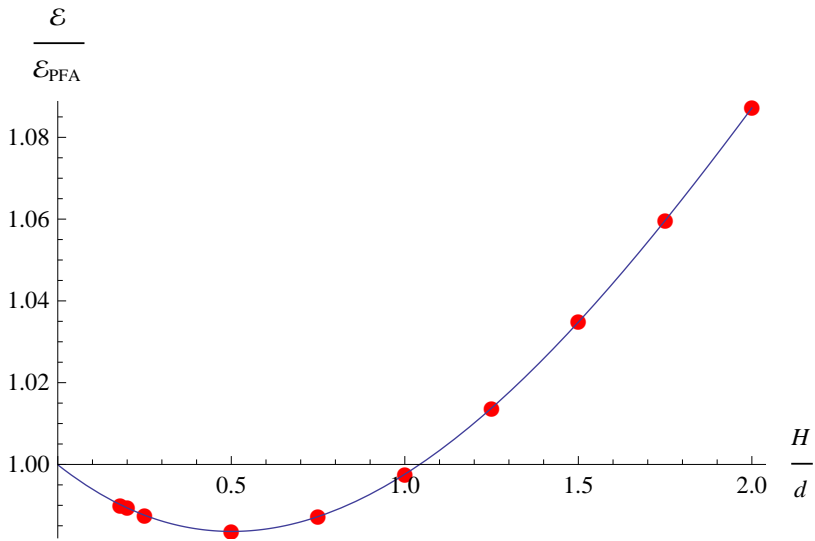}
\caption{Ratio of the exact Casimir energy to the proximity force
approximation for a perfectly conducting strip of width $2d$ parallel
to a perfectly conducting plane, as a function of separation $H$.  The
line represents a polynomial fit, from which we extract the
coefficients in Eq.~(\ref{eqn:expansion}).}
\label{fig:epfa}
\end{figure}
Results of the calculation for a strip parallel to a plane are shown
in Fig.~\ref{fig:epfa}.  We show the ratio of the full energy to the
proximity force approximation, where the latter is given by
\begin{equation}
\frac{\cal E_{\rm pfa}}{\hbar c L} = -\frac{\pi^2}{720} \frac{2d}{H^3}\,.
\end{equation}
We also show a polynomial fit to this quantity, which shows good
agreement with the expansion of Eq.~(\ref{eqn:expansion}), 
\begin{equation}
\frac{\cal E}{\cal E_{\rm pfa}}=
1 - \frac{2\beta H}{\frac{\pi^2}{720} 2d} 
- \frac{\gamma H^2}{\frac{\pi^2}{720} (2d)^2} + \ldots\,,
\end{equation}
from which we obtain $\beta=0.00092$ and
$\gamma=-0.0040$.  These dimensionless quantities give edge
corrections to the proximity force result:  $\beta$ captures the
effect of each of the two edges individually, while $\gamma$ gives an
interaction energy due to the combined effect of the two edges.  The
result for $\beta$ agrees with results obtained at lower precision in
the case of a half-plane parallel to a plane
\cite{wedge,parabolic1,parabolic2,planes}.  The strip allows for
better numerical precision than the half-plane, because the leading
proximity force term is a energy per unit length rather than an energy
per unit area.  (Of course, the subleading correction $\gamma$ is not
present in the half-plane case, since it has only a single edge.)

We can gain some qualitative insight into these corrections by
considering the effects of edges on the fluctuation modes that
contribute to the Casimir energy.  The positive sign of $\beta$
indicates that the edge boundary condition suppresses fluctuations
that would otherwise contribute to the attractive Casimir interaction
(though this effect arises only from terms beyond the first reflection
\cite{wedge,Babinet,planes}), while the negative sign of $\gamma$
indicates an enhancement of the Casimir energy due to the effect of
one edge on the other:  Some of the modes whose contribution would be
suppressed by one edge have already been suppressed by the other edge,
and so the combined effect of two edges reduces the Casimir energy by
less than the sum of their individual contributions.  We note that
Eq.~(\ref{eqn:energy}) is meromorphic around $H=0$, so we cannot have a
term proportional to $1/\ln (H/d)$ in Eq.~(\ref{eqn:expansion}).  In
contrast, for an  expansion at \emph{large} $H$, the essential
singularity in the integrand of Eq.~(\ref{eqn:energy}) makes it
possible for such inverse logarithms to appear, and indeed the leading
term at large distances is proportional to $1/[H^2 \ln (H/d)]$ in that
case \cite{Emig06}.

\section{Acknowledgments}

This work was supported in part by the National Science Foundation
(NSF) through grant PHY-1213456.
N.\ G.\ thanks G.\ Bimonte, T.\ Emig, R.\ L.\ Jaffe, M.\ Kardar, and
M.\ Kr\"uger for helpful conversations.

\bibliographystyle{apsrev}
\bibliography{article}

\end{document}